\documentclass[useAMS,usenatbib]{mn2e}
\usepackage{epsfig}
\usepackage{epstopdf}
\usepackage{lscape} 
\def\lesssim{\mathrel{\hbox{\rlap{\hbox{\lower4pt\hbox{$\sim$}}}\hbox{$<$}}}}

\def\gtrsim{\mathrel{\hbox{\rlap{\hbox{\lower4pt\hbox{$\sim$}}}\hbox{$>$}}}}

\title[picoarcsecond astrometry]{
50 pico arcsecond astrometry of pulsar emission
}
\author[Pen et al]{Ue-Li
  Pen$^{1}$\thanks{E-mail:\ pen@cita.utoronto.ca},
  J.P. Macquart$^2$\thanks{E-mail:\ jpmacquart@gmail.com},
  Adam Deller$^3$\thanks{E-mail:\ deller@astron.nl},
  Walter Brisken$^4$\thanks{E-mail:\ wbrisken@aoc.nrao.edu}
}
\begin{document}

\date{\today}

\pagerange{\pageref{firstpage}--\pageref{lastpage}} 
\pubyear{2012}

\maketitle
\label{firstpage}
\begin{abstract}

We use VLBI imaging of the interstellar scattering speckle pattern
associated with the pulsar PSR 0834+06 to measure the astrometric
motion of its emission.  The $\sim$5AU interstellar baselines, provided by
interference between speckles spanning the scattering disk, enable us
to detect motions with sub nanoarcsecond accuracy.
We measure a
small pulse deflection of $\sim 18 \pm 2$km (not including geometric
uncertainties), which is 100 times smaller than
the native resolution of this interstellar interferometer.  
This implies that the emission region is small, and at an altitude of
a few hundred km,
with the exact value depending on field geometry.  This is
substantially closer to the star than to the light cylinder.  Future
VLBI measurements can improve on this finding.  This new regime of
ultra-precise astrometry may enable precision parallax distance determination of
pulsar binary displacements.  

\end{abstract}

\newcommand{\be}{\begin{eqnarray}}
\newcommand{\ee}{\end{eqnarray}}
\newcommand{\beq}{\begin{equation}}
\newcommand{\eeq}{\end{equation}}

\section{Introduction}

The quantitative nature of pulsar emission has remained enigmatic for
the past half century since its discovery.  One limiting step has been
the lack of precise data.  To achieve sensivity to emission features
would require nano or pico arcsecond imaging, which is challenging to
achieve.  Various attempts have used the interstellar medium as a lens
to detect pulse emission motion: \cite{1987ApJ...320L..35W},
\cite{1999ApJ...520..173G}, \cite{2012ApJ...758....8J},
\cite{2012ApJ...758....7G}.  These have resulted in mutually difficult
to reconcile conclusions about the effective emission altitude.

More recently, \citet{2012MNRAS.421L.132P} suggested using VLBI
resolved images of the ISM scattering to coherently image the
pulsar. \cite{2010ApJ...708..232B} have mapped the individual
interstellar lensed images of B0834+06, with precise distances and
positions on the sky.  These plasma lensed images are separated by
many AU.  The ultimate objective of such a technique is to use these
individual lenses as apertures of a coherent interferometer; in
principle, one could reconstruct a full image if the emission region
is resolved An easier measurement is the relative phase change during
the pulsar rotation, which is presented in this letter.

\section{Data}

\begin{figure}
\centerline{\epsfig{file=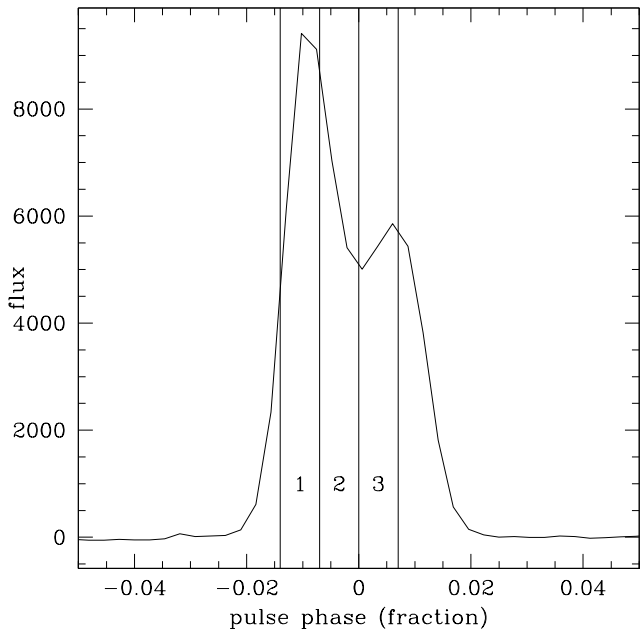,width=3.5in}}
\caption{
Correlation gates of 9ms width.  The 3 gates indicated in the plot are
used in this  analysis.  
}
\label{fig:gate}
\end{figure}

This analysis is based on a recorrelation of data used in
\cite{2010ApJ...708..232B}. The voltages of the Arecibo-GBT baselines
were recorrelated in gates of 8.9ms width, corresponding to 0.7\% of
the pulsar period.  Only the 3 gates shown in Figure \ref{fig:gate}
superimposed on the pulse
profile\footnote{http://www.naic.edu/\~{}pulsar/data/epndb/B0834+06/gl98\_408.epn.asc}
showed sufficient signal-to-noise for further processing -- the other
gates were not used further.

The pulsar and lens geometry are known: The pulsar is at a distance of
~640 pc (Deller \& Brisken, unpublished), and the screen is located a
distance of $415 \pm 5\,$pc (with this error estimate not including
pulsar distance uncertainties).  This places the screen two thirds of
the way to the pulsar, which improves the angular resolution: The
characteristic resolving power of the scattering is $\lambda/D \sim
0.3\,\mu$as for $\lambda \sim 1\,$m and a typical baseline length of
5\,AU, corresponding to displacement between major groups of speckles
on the scattering disk.  With the pulsar located 225\,pc from the
scattering region, this angular resolution corresponds to a linear
scale of 10,000\,km.  We used a wavelength $\lambda \sim 1$m and
$D=$5AU as a reference scale, which is roughly the displacement of a
major group of scattering points.  Two points at the pulsar separated
by 10,000 km differ by 2$\pi$ radians.  The native resolution, in this
sense, is then $\sim 1700$/SN km, where SN is the signal-to-noise of
the measurement.  From Earth, this corresponds to a resolution of
100/SN nano arcseconds.

For this analysis, we primarily used data acquired on the Arecibo (AO)
-- Greenbank (GB) interferometric baseline.  The individual dish data
was difficult to interpret due to finite bit sampling and local RFI.
The only other data used in this analysis was the AO-Jodrell Bank 76m
(JB) cross-correlation baseline, needed for the screen image in
Section 3.  For this baseline, a single gate of width 125 ms spanning
the entire on-pulse (taken from the original
\cite{2010ApJ...708..232B} visibility data) was used, since the raw JB
voltage data was no longer available for recorrelation.

\section{Holography}
\label{sec:holo}

Our objective is to perform relative astrometry on the pulsar as a
function of pulse phase.  To increase the sensitivity of our
astrometry, we applied a holographic technique to obtain the phase of
the pulsar radiation through each scattering image.  We used a
technique similar to \cite{2008MNRAS.388.1214W}.  The basic framework
is the same, which we summarize briefly.  The dynamic spectrum
$I(\nu,t) = \langle v(\nu,t') \bar{v}(\nu,t')\rangle_{t'\sim t}$ is
the average modulus of the complex field response.  

We call the Fourier transform of the dynamic spectrum, $\tilde I
(\tau, \omega)$, the conjugate spectrum.  This quantity is the auto
correlation of the fourier transform of the voltage impulse-response
function.  \cite{2008MNRAS.388.1214W} showed that the voltage
impulse-response function decomposes into a sparse set in the
delay--Doppler rate ($\tau-\omega$) plane.  Its modulus is the
secondary spectrum, which is a quartic quantity in the
impulse-response function.

\begin{figure}
\centerline{\epsfig{file=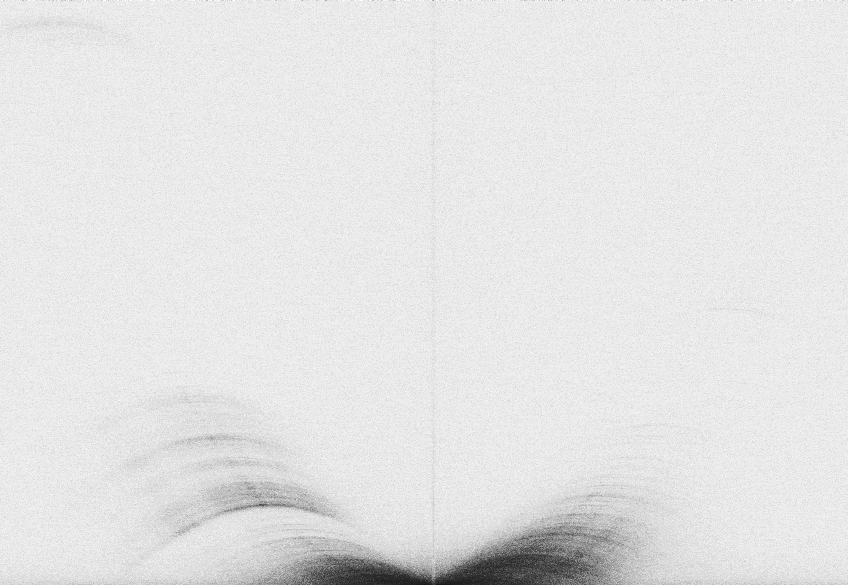,width=2.83in}}
\vspace{-3.13in}
\centerline{\hspace{-0.079in}\epsfig{file=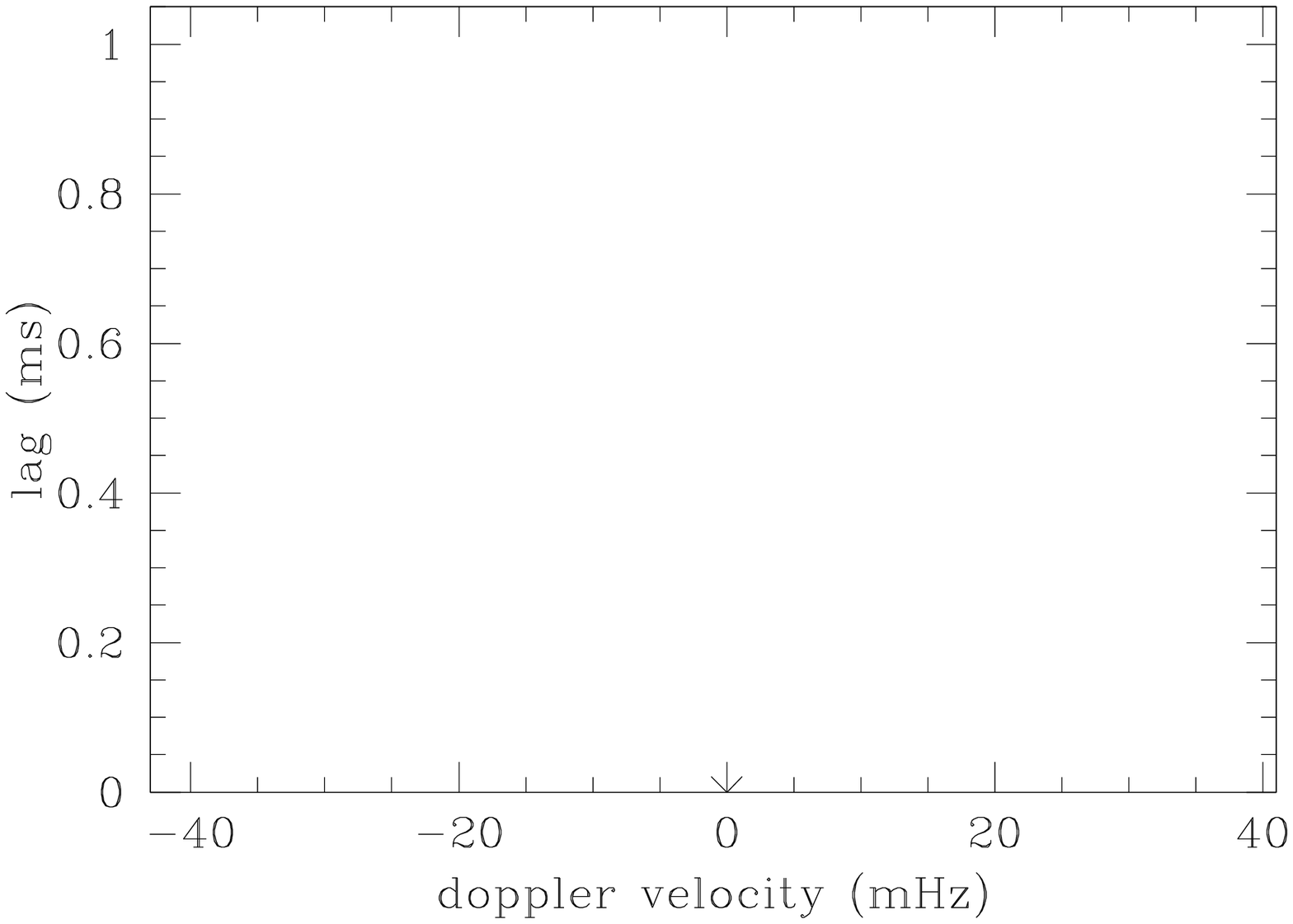,width=3.5in}}
\caption{
Secondary spectrum. The horizontal axis is $\pm 40$ mHz,
the vertical axis extends to 1ms.
}
\label{fig:ss}
\end{figure}

We show the secondary spectrum summed over all pulse gates in Figure 
\ref{fig:ss}.  As originally noted by \cite{2001ApJ...549L..97S}, the
striking feature is the presence of inverted parabolic arclets.

In the conjugate spectrum, the modulus is an auto-correlation of the
voltage impulse-response function.  An isolated inverted arclet
corresponds to the correlation of an isolated scattering point with
the main scattering disk.  By convolving the conjugate spectrum with
the complex conjugate of an inverted arclet, we obtain the analogy of
a ``dirty image'' of the holographic impulse-response function.

\begin{figure}
\centerline{\epsfig{file=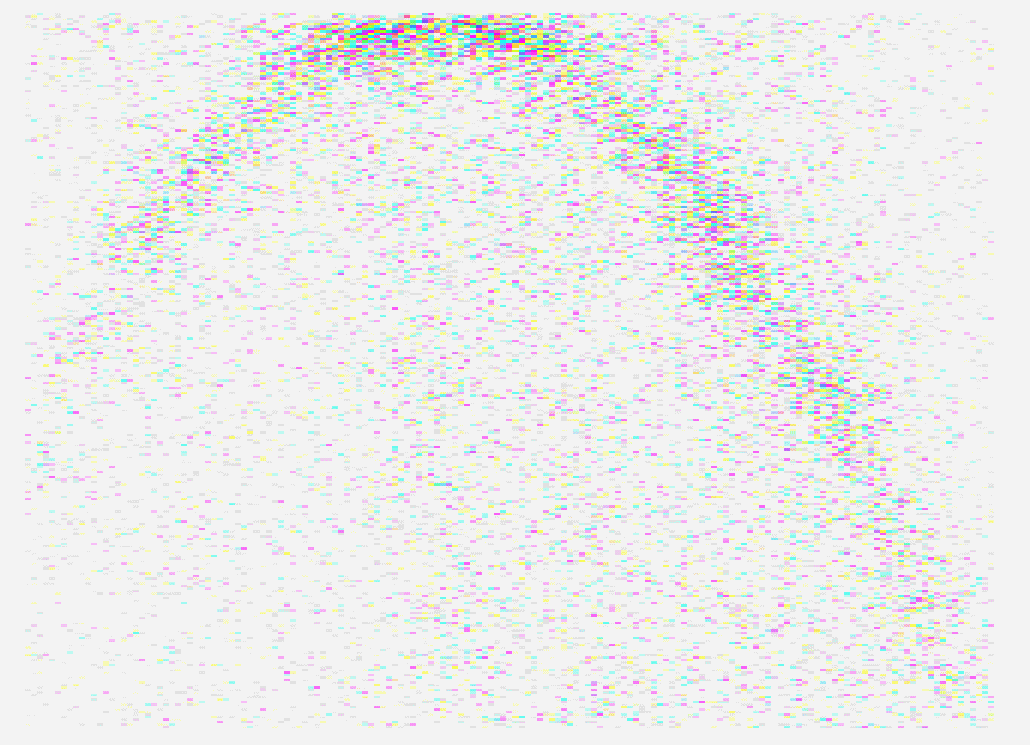,width=3.5in}}
\caption{
coherently stacked arclets.  The colour coding represents the phase.
It appears random, but is essential to achieve coherent holography. 
}
\label{fig:sa}
\end{figure}

Unlike the Walker procedure, we do not start with the brightest points
in the secondary spectrum.  These regions contain contributions from
many scattering points, and are maximally degenerate on their own.
Instead of decomposing the regions of brightest flux, we start with an
isolated inverted arclet in the conjugate spectrum.  This arclet is
primarily the impulse-response function of the voltage as measured against an
isolated scattering point, and allows a solution of the impulse-response
function local in the secondary spectrum.  We first isolate one, and
convolve the conjugate spectrum by it.  We then pick an isolated
region of bright response points around -20mHz, and stack the arclets
weighted by the conjugate of these response points.  Figure
\ref{fig:sa} shows a stack of arclets.  These are complex impulse-response
functions.

\begin{figure}
\centerline{\epsfig{file=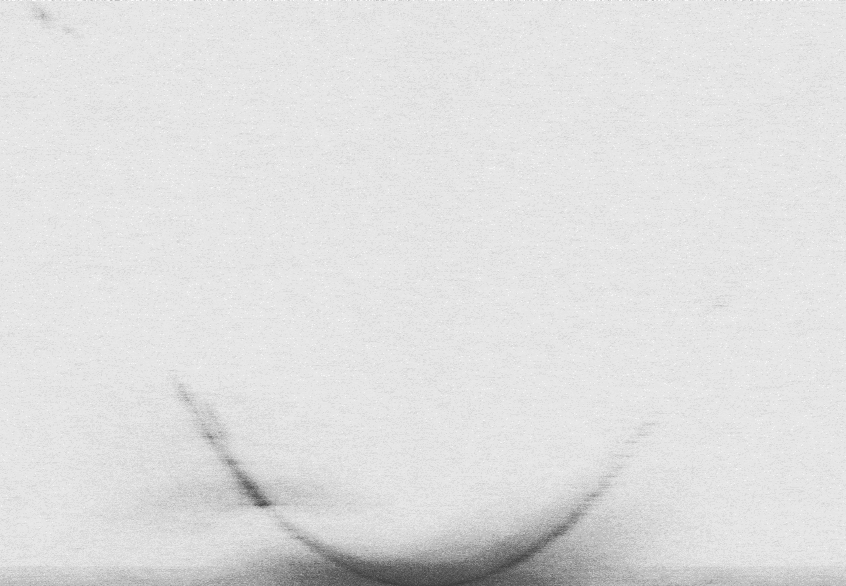,width=2.83in}}
\vspace{-3.13in}
\centerline{\hspace{-0.079in}\epsfig{file=ssgrid.eps,width=3.5in}}
\caption{
Holographic gate cross spectrum obtained by multiplying the semi
descattered conjugate
spectrum of gate 1 with the complex conjugate of the corresponding
quantity of gate 2.  Shown here is the real part.
}
\label{fig:ha}
\end{figure}

The conjugate spectrum is then again convolved, this time with the
stacked arclet template.  We call this the dirty semi-descattered
spectrum, indicating that the complex impulse-response function has
been applied once. A cleaned fully descattered spectrum would be a
dynamic spectrum which does not scintillate in frequency or time.  In
principle, this dirty semi-descattered spectrum maximizes the
signal-to-noise on each residual in the limit that individual
scattering points remain uncorrelated. The result is shown in Figure
\ref{fig:ha}.

We see the inverted parabolic arclets descattered onto the main
parabola. While the stacked arclets are obtained from the combined
gates, we apply this to each gate separately.  Then we multiply the
conjugate spectrum from one gate by the complex conjugate of the
conjugate spectrum from another gate.  If these two are statistically
identical, the expectation value will be a real number.  If there is a
systematic phase lag, the imaginary component will not be zero.
Initial examination shows the imaginary part of the holographic cross
spectrum to be much smaller than the real part, indicating that
phase changes are very small.  We expect the phase change to be
proportionate to the linear separation between images, which is equal
to the doppler frequency for this collinear scattering structure.

To estimate the noise in the holographic cross spectrum, we apply a
bootstrap technique: for each pair of gates, we generate new gates by
randomly selecting scintels from each constituent dynamic spectrum.  We
then compute the holographic cross spectrum.  In this construction,
the imaginary part has by construction zero expectation value and a
finite variance.  We compute the variance at each point from 1000
realizations.  Knowing the variance allows us to fit for the phase
gradient, and estimate its error.  We examined the covariance between
bins, which is less than 10\% for neighboring bins, and consistent
with zero for further separated bins.  We thus neglect the covariance,
since an inverse of the noise matrix would be very noisy and require
many more realizations.

\begin{figure}
\vspace{0.5in}
\centerline{\epsfig{file=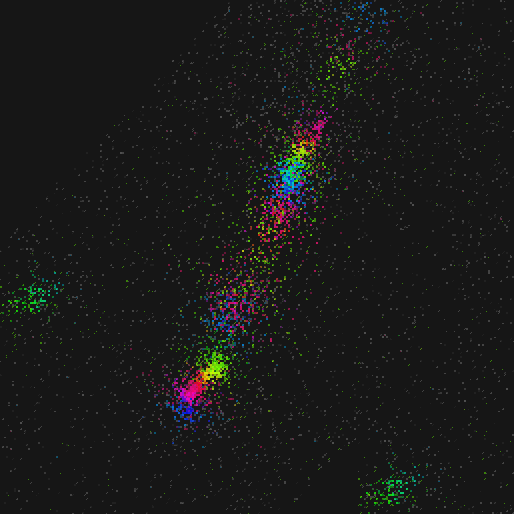,width=2.83in}}
\vspace{-3.13in}
\centerline{\hspace{-0.079in}\epsfig{file=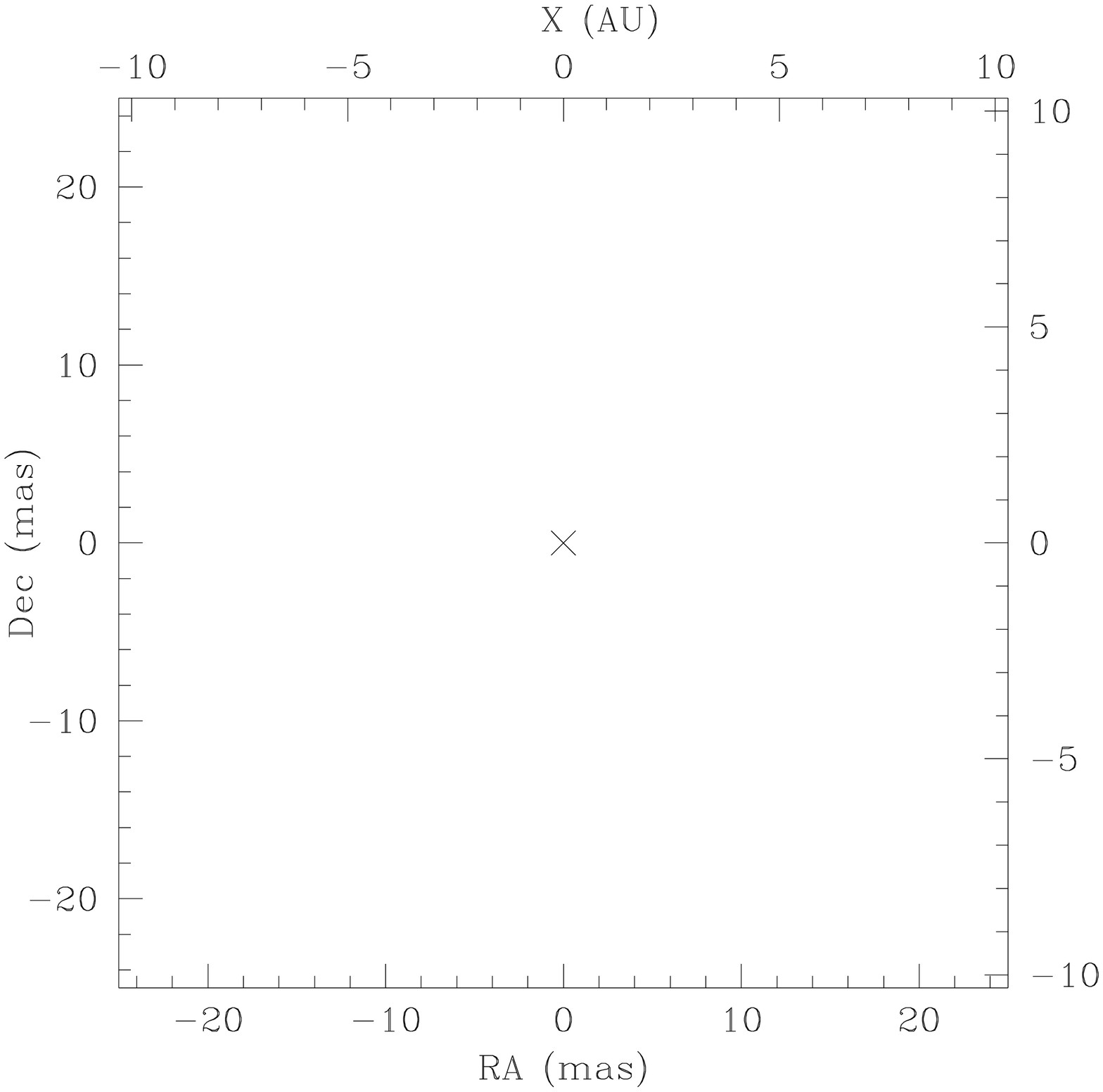,width=3.5in}}
\caption{
Holographic VLBI image. Imaged formed using the VLBI
phase of each holographic cross spectrum residual. The colour is
proportionate to time delay, in a periodic hue map.  The green locus
of points at left center is a sidelobe of the true locus on the lower
right. The main base of the parabola as been omitted, which is
centered at the origin.  This scattering image becomes our celestial
interferometer aperture with which we image the pulsar.  At a
scattering screen distance of $\sim$400 pc, each mas is 0.4 AU.
}
\label{fig:vlbi}
\end{figure}

We also used the holographic technique to directly image the pulsar
scattering image on the sky.  The same procedure results in a
semi-descattered VLBI visibility and secondary cross spectrum
\citep{2010ApJ...708..232B}, i.e. a complex value for each doppler
frequency and lag.  The phase determines the distance along the
baseline.  We used only the AO-GBT and AO-JB baselines, and
interpreted each phase as the distance along the baseline.  This is
then rebinned onto a 2-D map of the sky, shown in Figure
\ref{fig:vlbi}

The map shows a visual representation of the scattering disk obtained
this way.  This forms the celestial interferometer which we use for
astrometry of the pulse emission.  This is the aperture of our
interstellar interferometer which we will use for precision imaging of
the pulsar.

\section{Results}

\begin{figure}
\centerline{\epsfig{file=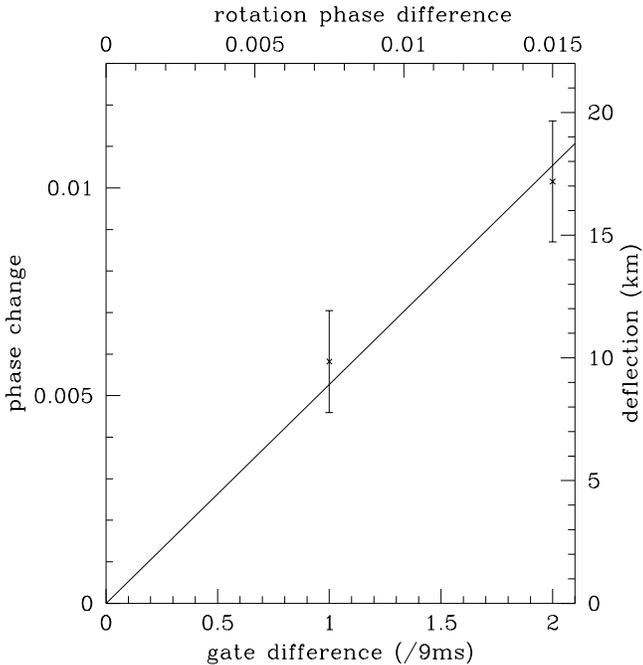,width=3.5in}}
\caption{Pulsar motion projected along scattering axis.  The
  horizontal axis is the time difference, 
  the vertical axis is the apparent motion shift along the scattering
  axis, scaled to a doppler frequency of -15mHz.  The changes are
  always small.  The error is about 1/1000th of a radian, wich reflects the
extremely high  signal-to-noise of the measurement.}
\label{fig:angle}
\end{figure}

In Figure \ref{fig:angle} we show the reconstructed pulse motion. The
first feature to notice is that the motion is tiny: different gates
have almost identical phase responses on the scattering screen to
within one percent of a radian.  This immediately implies that the
motion is less than 30km.

This is two orders of magnitude smaller than some previous
estimates\citep{1987ApJ...320L..35W,1999ApJ...520..173G}, albeit for
different pulsars, and surprisingly small compared to some size
interpretations using the modulation index\citep{2012ApJ...758....7G}.
Recent modulation analyses have also reported sizes consistent with
zero, and smaller than 4km\citep{2012ApJ...758....8J}.  We note that
our measurement is differential, relying on only the change of
scintillation pattern from gate to gate.

We find the best linear motion fit to be $9\pm 1.1$ km per gate of
9ms, which is an apparent motion of $\sim$ 1000 km/sec.  The pulsar
motion is $\sim 162\pm 10$ km/sec\citep{1982MNRAS.201..503L} due
north, which projects to $\sim 147$ km/sec along the scattering structure,
almost an order of magnitude smaller than the measured motion.  We do
not correct for this effect.

No motion is observed on the 1ms island, which is at about 45 degrees
to the main scattering axis.  This island has about 4\% of the flux,
limiting the motion of the pulse along the dimension transverse to the
scattering disk to $\lesssim 35$ km ($2-\sigma$).

\section{Interpretation}

The direct observable is the apparent motion of the emission region as
a function of pulse phase.  This can in turn be related to the pulsar
emission altitude by specifying the geometry of the emission
mechanism.  We consider some limits that are exactly calculable.  One
is the case that the emission region is very small, and the pulse
period is determined by the beaming angle of the emission.  In this
case, the deflection of the pulse is just the height times the duty
cycle, corresponding to an emission height of about 200km.

Possibly the extreme antithetical model is one in which the emission
is infinitely beamed radiating tangentially to the local magnetic
field line. In this case, the altitude is related to the deflection by
the field curvature.  If we take the curvature to be roughly the
altitude, as expected for a dipolar field, we obtain the same rough
altitude as above.  We note that this can be arbitrarily incorrect: in
the unlikely extreme case that the field lines are purely radial with
no curvature, there is no apparent motion, and the emission altitude
could be arbitrarily high.  In the infinite curvature limit, the
emission is given by the same altitude as the beamed case.

From these scenarios we also see that the emission region generically
has a smaller angular size than the displacement during the pulse: in
the infinitely beamed scenario we only see the one field line
tangential to our line of sight, resulting in an infinitesimal
apparent emission size.  In the first scenario, we see all field lines
which are within one beaming angle of our line of sight.  The size of
the emission region plus the beaming angle must be less than the pulse
duration, while the deflection is given by the beaming angle.  There
could be exotic counter examples, for example if the field lines were
convergent at the emission region.

\section{Conclusions}

We have used the interstellar medium is a giant lens to study the
motion of the pulsar emission.  We have found this motion to be tiny,
by more than order of magnitude, compared to previous indirect
indications.  We expect the apparent size of the emission region to be
smaller than the displacement, making the emission essentially
unresolvable on these interstellar baselines.  More VLBI observations
of a range of pulsars at lower frequencies could help improve the
situation.

The small nature of the emission region implies that pulsars will be
unresolved by all known scattering screens, including the scattering
screen at the galactic center\citep{2009ApJ...702L.177D}, and
potentially multiple images against a black
hole\citep{2006PhRvD..73f3003R}.

This new era of precision picoarcsecond astrometry opens the
possibilities for precise distance measurements to pulsars.  It would
increase the sensitivity and angular resolution of pulsar timing
arrays to gravitational waves\citep{2012PhRvD..86l4028B}.  The pulsar
reflex motion for a binary system provides the appropriate parallax.

\section{Acknowledgements}

U-LP thanks NSERC and CAASTRO for support.

\newcommand{\araa}{ARA\&A}   
\newcommand{\afz}{Afz}       
\newcommand{\aj}{AJ}         
\newcommand{\azh}{AZh}       
\newcommand{\aaa}{A\&A}      
\newcommand{\aas}{A\&AS}     
\newcommand{\aar}{A\&AR}     
\newcommand{\apj}{ApJ}       
\newcommand{\apjs}{ApJS}     
\newcommand{\apjl}{ApJ}      
\newcommand{\apss}{Ap\&SS}   
\newcommand{\baas}{BAAS}     
\newcommand{\jaa}{JA\&A}     
\newcommand{\mnras}{MNRAS}   
\newcommand{\nat}{Nat}       
\newcommand{\pasj}{PASJ}     
\newcommand{\pasp}{PASP}     
\newcommand{\paspc}{PASPC}   
\newcommand{\qjras}{QJRAS}   
\newcommand{\sci}{Sci}       
\newcommand{\solphys}{Solar Physics}       %
\newcommand{\sova}{SvA}      
\newcommand{\aap}{A\&A}

\bibliography{pulse}
\bibliographystyle{mn2e}

\label{lastpage}

\end{document}